\documentclass[aps,print]{revtex4}
\usepackage{amsfonts}
\usepackage{amsmath}
\usepackage{amssymb}
\usepackage{graphicx}
\usepackage{color}

\begin{document}

\title{Vortex-ring quantum droplets in a radially-periodic potential}
\author{Bin Liu$^{1,2}$, Yi xi Chen$^{1}$, Ao wei Yang$^{1}$, Xiao yan Cai$%
^{1}$, Yan Liu$^{3}$, Zhi huan Luo$^{3}$, Xi zhou Qin$^{1,2}$, Xun da Jiang$%
^{1,2}$}
\email{jxd194911@163.com}
\author{Yong yao Li$^{1,2}$}
\author{Boris A. Malomed$^{4,5}$}
\affiliation{$^{1}$School of Physics and Optoelectronic Engineering, Foshan University,
Foshan 528000, China}
\affiliation{$^{2}$Guangdong-Hong Kong-Macao Joint Laboratory for Intelligent Micro-Nano
Optoelectronic Technology, Foshan University, Foshan 528000, China}
\affiliation{$^{3}$Department of Applied Physics, College of Electronic Engineering,
South China Agricultural University, Guangzhou 510642, China}
\affiliation{$^{4}$Department of Physical Electronics, School of Electrical Engineering,
Faculty of Engineering, and Center for Light-Matter Interaction, Tel Aviv
University, Tel Aviv 69978, Israel}
\affiliation{$^{5}$Instituto de Alta Investigaci\'{o}n, Universidad de Tarapac\'{a},
Casilla 7D, Arica, Chile}

\begin{abstract}
We establish stability and characteristics of two-dimensional (2D) vortex
ring-shaped quantum droplets (QDs) formed by binary Bose-Einstein
condensates (BECs). The system is modeled by the Gross-Pitaevskii (GP)
equation with the cubic term multiplied by a logarithmic factor (as produced
by the Lee-Huang-Yang correction to the mean-field theory) and a potential
which is a periodic function of the radial coordinate. Narrow vortex rings
with high values of the topological charge, trapped in particular circular
troughs of the radial potential, are produced. These results suggest an
experimentally relevant method for the creation of vortical QDs (thus far,
only zero-vorticity ones have been reported). The 2D GP equation for the
narrow rings is approximately reduced to the 1D form, which makes it
possible to study the modulational stability of the rings against azimuthal
perturbations. Full stability areas are delineated for these modes. The
trapping capacity of the circular troughs is identified for the vortex rings
with different winding numbers (WNs). Stable compound states in the form of
mutually nested concentric multiple rings are constructed too, including
ones with opposite signs of the WNs. Other robust compound states combine a
modulationally stable narrow ring in one circular potential trough and an
azimuthal soliton performing orbital motion in an adjacent one. The results
may be used to design a device employing coexisting ring-shaped modes with
different WNs for data storage.
\end{abstract}

\maketitle

\section{\qquad Introduction}

Quantum droplets (QDs) were predicted, as a new species of quantum matter,
by Petrov and Astrakharchik \cite{Petrov2015,Petrov2016}, and then promptly
created in dipolar \cite{Schmitt2016,Chomaz2016} and binary bosonic gases,
both homoatomic \cite{Cabrera2018,Leticia,Inguscio} and heteroatomic \cite%
{Chiara} ones. The QDs are maintained by the balance of the effective
mean-field attraction (which itself is a result of the competition of the
inter-component attraction and intra-component self-repulsion), driving the
system towards the collapse, and the higher-order self-repulsion in each
component induced by quantum fluctuations, which is represented by the
Lee-Huang-Yang (LHY) correction \cite{LHY} to the respective
Gross-Pitaevskii (GP) equations. This correction takes different forms in
3D, 2D and 1D versions of the GP equations (in particular, its sign flips
into effective self-attraction in 1D) \cite{Petrov2016,Ilg,PRA103_013312}.

QDs have drawn much interest as stable self-trapped modes, both
multidimensional and one-dimensional
\cite{ PRA101_051601, PRA98_033612,PRA102_043302, PRResearch4_013168,PRA102_023318, PRL126_025301,PRResearch2_033522,PRE102_062217,FOP16_32201_LYY,PRA99_053602,CSF152_111313,CXL_PRA2018,PRL120_160402,CNSNS_LZD,PRL126_025302,PRA104_043301,PRA103_033312,PRResearch2_043074,PRL126_244101, PRResearch3_033247,CSF_WHC,FOP_GMY,FOP_Boris,CSF2022_ZJF,wlxb_cyx,Nonlinear Dyn. Zhou 2022,Front. Phys. Hu 2022,CSF Zhou 2021,NJP1,NJP2,NJP3,NJP4,NJP5,NJP6,NJP7,NJP8,NJP9,BA1,BA2,book}.%
In addition to their significance to
fundamental studies, they offer potential applications, such as the design
of matter-wave interferometers \cite{nature539_176}.

A remarkable result is the prediction of the fact that QDs with embedded
vorticity (topological charge, alias winding number, WN), $S\geq 1$, may
also be stable, in the 2D \cite{PRA_LYY_2DQD2018} and 3D \cite{KTS2018}
settings alike. These are essential findings, as the stability is a major
issue for fundamental and vortical multidimensional soliton-like states \cite%
{PhysD,book}. Stable semidiscrete vortex QDs were also predicted in arrays
of coupled 1D cigar-shaped traps \cite{PRL123_133901}.

However, the creation of vortex QDs in an experiment has not been reported
yet, and it is considered as a technically challenging objective. Therefore,
search for settings which facilitate the existence of such stable
topological modes is a relevant problem. A known possibility for that is
offered by the use of spatially periodic \cite{KonBra,BEC_OL} or
quasiperiodic \cite{quasi} lattice potentials. It was predicted that they
not only stabilize vortex modes with various degrees of complexity \cite%
{BBB,Ziad,HS}, but also alter formation patterns in many ways \cite%
{ZHF_CPB,PRA71_053611,ZX_JPB,PRA96_043626,CPL30_060306,AML92_15,ActaPS63_190502,annals of p322_1961,CPB21_020306, pra99_023630,ActaPS68_043703,PRA95_043618,mplb32_1850070,PRA98_033827,pre74_066615,PRA64_063608,DLW_RP,HYJ_RP,YFF_RP,anti, 1D_AVK_CQ,checkerboard_potential,toroidal_ZYP,toroidal_PRL99_260401,toroidal_PRA99_043613, toroidal_PRA84_063638,PRA95_033606}. In 1D settings with periodic potentials,
QDs were studied too \cite{ZZ_CNSNS78_104811,ND_DLW102_303}.\ Vortex QDs
with $S=1$ in the 2D setting including the square-lattice potential were
considered in Ref. \cite{FOP_ZYY}. On the other hand, because the square
lattice breaks the rotational symmetry and conservation of the angular
momentum, it is more natural to use radial (axisymmetric) lattices as an
experimentally relevant means supporting stable self-trapped vortex modes,
especially ones with high WN\ values. In particular, Ref. \cite{PRA_HCQ2017}
reported stable vortex solitons with $S=11$ maintained by the radial lattice
in the dipolar Bose-Einstein condensate (BEC) with repulsive long-range
dipole-dipole interactions.

The objective of this work is to demonstrate that stable 2D vortex QDs,
which are formed by binary BECs with contact (rather than long-range)
interactions, can be effectively made and controlled with the help of a
radially periodic axisymmetric potential, which directly suggests a new
approach for the creation of vortex QDs in experiments. The system is
modeled by the GP equation with this potential and LHY term. The respective
vortex QDs are ring-shaped, being trapped in particular circular troughs of
the radial potential with multiple minima, cf. Refs. \cite%
{Vyslo1,Vyslo2,pre74_066615}. The difference from the above-mentioned stable
self-trapped vortex states in the free space is that those ones may not be
stable unless they are very broad \cite{PRA_LYY_2DQD2018} (which is an
obvious difficulty for the creation of such modes in the experiment), while
it is shown below that the axisymmetric lattice potential may support narrow
vortex rings as stable modes, thus offering an advantage for the
experimental realization. For narrow rings, the 2D GP equation is
approximately reduced to its 1D form by eliminating the radial coordinate.
The modulational stability of the rings in the azimuthal direction is then
analyzed in the framework of the 1D equation. Effects of the depth and
period of the radial potential on the vortex ring-shaped QDs in the full 2D
form are systematically studied by means of numerical methods, and stability
areas are identified for them. Composite vortex patterns, built of
concentric rings nested in different circular troughs, which carry different
WNs, including the case when a vortex ring is embedded in a larger one with
the opposite sign of the WN, are considered too (while stationary multi-ring
patterns do not exists in the free space). Earlier, multi-ring concentric
patterns with different WNs in different rings were introduced in Ref. \cite%
{Skarka} as solutions to the 2D complex Ginzburg-Landau equation with the
cubic-quintic nonlinearity (without the radial potential); however, unlike
the present setting, those pattern are dissipative ones. These results offer
a possibility to design a new encoding device, using the coexisting rings
with different WNs for storing different data components.

The rest of the paper is structured as follows. The model is introduced in
Sec. II, and analytical results produced by the reduced 1D GP\ equation are
presented in Sec. III. Numerical findings for the 2D vortex ring-shaped QDs
are summarized in Sec. IV, and the multi-ring vortex QDs combining different
WNs in concentric circular troughs are the subject of Sec. V. The work is
concluded by Sec. VI.

\section{The model}

According to Ref. \cite{FOP_ZYY,PRA_LYY_2DQD2018}, we assume that QDs, which
are formed by the binary BEC, are strongly confined in the transverse
direction, having a lateral size $l\gg \sqrt{a_{1,2}a_{\bot }}$, where $%
a_{1,2}$ and $a_{\bot }$ are the scattering lengths accounting for the
self-repulsion of each component and the transverse confinement length,
respectively. This condition can be readily implemented in a typical
experimental setup, with $a_{1,2}\sim 3$ nm, $a_{\bot }\lesssim 1$ $\mathrm{%
\mu }$m, and $l\sim 10$ $\mathrm{\mu }$m. Assuming the symmetry between the
two components, i.e., $a_{1}=a_{2}$, the system of coupled GP equations
including the LHY correction is reduced to the scaled 2D form \cite%
{Petrov2016},%
\begin{equation}
i\frac{\partial }{\partial t}\Psi _{j}=\left( -\frac{1}{2}\nabla ^{2}+V(r)+%
\frac{\delta E_{\mathrm{2D}}}{\delta n_{j}}\right) \Psi _{j},  \label{GPE}
\end{equation}%
where $j=1,2$ are labels for the two species with wave functions $\Psi _{j}$
and densities $n_{j}=|\Psi _{j}|^{2}$, $\nabla ^{2}=\partial
_{r}^{2}+r^{-1}\partial _{r}+r^{-2}\partial _{\theta }^{2}$ is the Laplacian
written in the polar coordinates, and $V(r)$ is the radial potential. The
total potential energy $E_{\mathrm{2D}}$ of the system includes the
mean-field and LHY terms, \textit{viz}.,
\begin{equation}
E_{\mathrm{2D}}=\frac{1}{2}\int_{0}^{\infty }rdr\int_{0}^{2\pi }d\theta %
\left[ g(n_{1}-n_{2})^{2}+(n_{1}+n_{2})^{2}\ln \left( \frac{n_{1}+n_{2}}{%
\sqrt{e}}\right) \right] ,  \label{totalenergy}
\end{equation}%
where $g$ is the mean-field coupling constant \cite{Petrov2016}. In the
symmetric state, with $\Psi _{1}=\Psi _{2}\equiv \Psi /\sqrt{2}$ and $%
n_{1}=n_{2}=n/2\equiv \left\vert \Psi ^{2}\right\vert /2$, the system of
equations (\ref{GPE}) amounts to a single LHY-corrected GP equation:
\begin{equation}
i\frac{\partial \Psi }{\partial t}=-\frac{1}{2}\nabla ^{2}\Psi +V\left(
r\right) \Psi +\ln \left( \left\vert \Psi \right\vert ^{2}\right) \cdot
\left\vert \Psi \right\vert ^{2}\Psi .  \label{GPE1}
\end{equation}%
Note that the nonlinearity with the logarithmic factor in Eq. (\ref{GPE1}), $%
\ln n$, implies self-attraction at $n<1$ and self-repulsion at $n>1,$
respectively.

\begin{figure}[tbp]
{\includegraphics[width=0.7\columnwidth]{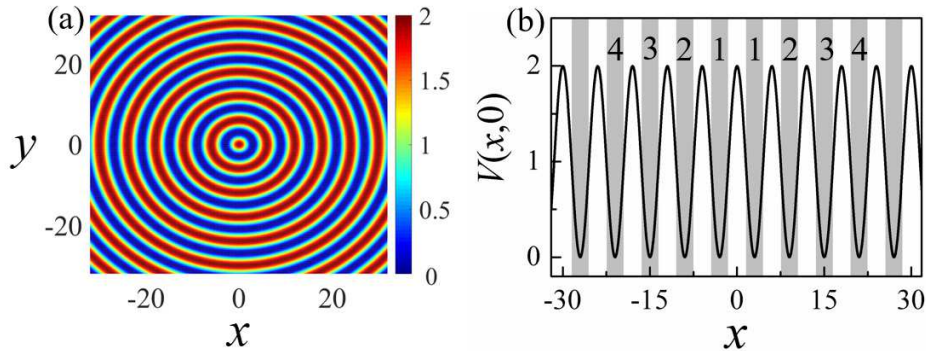}}
\caption{(Color online) (a) The radially periodically potential\ (\protect
\ref{potential}) with $V_{0}=2$. (b) Its cross section, $V(x,0)$, along $y=0$%
. In the latter panel, numbers $1$, $2$, $3$, ... represent integer values O$%
_{\text{n}}$ in Eq. (\protect\ref{RR}).}
\label{Fig_potential}
\end{figure}

Similar to Ref. \cite{pre74_066615}, we choose the axisymmetric potential
with depth $V_{0}>0$ (so that the origin at $r=0$ represents a local maximum
of the potential) and radial period $D$ (with experimentally relevant values
$\sim $ a few $\mathrm{\mu }$m),
\begin{equation}
V\left( r\right) =V_{0}\cos ^{2}\left( \frac{\pi }{D}r\right) ,
\label{potential}
\end{equation}%
see Fig. \ref{Fig_potential}(a). This potential can be created by a
cylindrical laser beam passed through an appropriate amplitude-modulation
plate. An essential property of the potential with multiple concentric
troughs is that ring-shaped states can self-trap in particular troughs. They
are characterized by the total norm,
\begin{equation}
N=\int_{0}^{+\infty }rdr\int_{0}^{2\pi }d\theta \left\vert \Psi \left(
r,\theta \right) \right\vert ^{2}.  \label{N}
\end{equation}

We look for stationary solutions to Eq. (\ref{GPE1}) which represent vortex
QDs with chemical potential $\mu $ and integer WN $S$ as
\begin{equation}
\Psi \left( r,\theta ,t\right) =\psi \left( r\right) \exp \left( -i\mu
t+iS\theta \right) ,  \label{Solution}
\end{equation}%
where $\psi \left( r\right) $ is a real radial wave function satisfying the
equation%
\begin{equation}
\mu \psi =-\frac{1}{2}\left( \frac{d^{2}\psi }{dr^{2}}+\frac{1}{r}\frac{%
d\psi }{dr}-\frac{S^{2}}{r^{2}}\psi \right) +V_{0}\cos ^{2}\left( \frac{\pi
}{D}r\right) \psi +\ln \left( \psi ^{2}\right) \cdot \psi ^{3}.
\label{radial}
\end{equation}%
Stationary vortex ring-shaped QDs, trapped in a radial potential trough
labeled by number O$_{\text{n}}=1,2,3...$, as shown in Fig. \ref%
{Fig_potential}(b), were produced by means of the imaginary-time-integration
method \cite{ITP1,ITP2}. It was applied to an initial guess taken as

\begin{equation}
\Psi _{0}\left( r,\theta \right) =C\exp \left[ -\alpha \left( r-r_{\text{n}%
}\right) ^{2}+iS\theta \right] ,  \label{initial-guess}
\end{equation}%
where $C$ and $\alpha >0$ are real numbers, and
\begin{equation}
r_{\text{n}}=\left( \text{O}_{\text{n}}-1/2\right) D,  \label{RR}
\end{equation}%
is the radial coordinate of the trough's bottom point. Then, the stability
of the stationary QDs was analyzed by means of the linearized equations for
perturbations with infinitesimal amplitude $\varepsilon $, which are
introduced as%
\begin{equation}
\Psi \left( r,\theta ,t\right) =\left\{ \psi (r)+\varepsilon \left[
u(r)e^{\gamma t+im\theta }+v^{\ast }(r)e^{\gamma ^{\ast }t-im\theta }\right]
\right\} e^{-i\mu t+iS\theta },  \label{pertur_solution}
\end{equation}%
where $u(r)$, $v(r)$, and $\gamma $ are eigenmodes and the instability
growth rate corresponding to an integer azimuthal index $m$ of the
perturbation, and $\ast $ stands for the complex conjugate. The
linearization around the stationary solution leads to the system of the
Bogoliubov -- de Gennes equations,

\begin{equation}
\left(
\begin{array}{cc}
F_{11} & F_{12} \\
F_{21} & F_{22}%
\end{array}%
\right) \left(
\begin{array}{c}
u \\
v%
\end{array}%
\right) =i\gamma \left(
\begin{array}{c}
u \\
v%
\end{array}%
\right) ,  \label{BDG_Eq}
\end{equation}%
where
\begin{align*}
F_{11}& =-\frac{1}{2}\left[ \frac{d^{2}}{dr^{2}}+\frac{d}{rdr}-{\frac{\left(
S+m\right) ^{2}}{r^{2}}}\right] +2\psi ^{2}\ln \left( \psi ^{2}\right) +\psi
^{2}+V(r)-\mu , \\
F_{12}& =\psi ^{2}\ln \left( \psi ^{2}\right) +\psi ^{2}, \\
F_{21}& =-\psi ^{2}\ln \left( \psi ^{2}\right) -\psi ^{2}, \\
F_{22}& =\frac{1}{2}\left[ \frac{d^{2}}{dr^{2}}+\frac{1}{r}\frac{d}{dr}-{%
\frac{\left( S-m\right) ^{2}}{r^{2}}}\right] -2\psi ^{2}\ln \left( \psi
^{2}\right) -\psi ^{2}-V(r)-\mu .
\end{align*}

Numerical solutions of Eq. (\ref{BDG_Eq}), obtained by means of the
finite-difference method, produced a spectrum of eigenvalues $\gamma $, the
stability condition being that the entire spectrum must be pure imaginary
\cite{Mihalache,Nir}. Then, the stability of the stationary solutions was
verified by direct simulations of the perturbed evolution in the framework
of Eq. (\ref{GPE1}). The simulations were performed by means of the {fast
Fourier-transform method}.

\section{Analytical results: modulational (in)stability of narrow vortex
rings and quasi-1D solitons}

The fact that the 2D GP equation (\ref{GPE1}) gives rise to narrow
vortex-ring modes populating particular circular troughs (see Fig. \ref%
{example} below) suggests to approximate it by the 1D equation with the
angular coordinate $\theta $, while the radial coordinate $r$ is eliminated.
To this end, we adopt an approximate ansatz%
\begin{equation}
\Psi (r,\theta ,t)=\psi _{0}\exp \left[ -i\left( \chi +\frac{S^{2}}{2r_{%
\text{n}}^{2}}\right) t+iS\theta -\frac{\left( r-r_{\text{n}}\right) ^{2}}{%
2w^{2}}\right] \psi \left( \tilde{\theta},t\right) ,  \label{ansatz}
\end{equation}%
with the 1D wave function $\psi (\tilde{\theta},t)$, which is defined with
respect to the rotating angular coordinate,
\begin{equation}
\tilde{\theta}=\theta -\left( S/r_{\text{n}}^{2}\right) t.  \label{tilde}
\end{equation}%
The radial center of ansatz (\ref{ansatz}) is pinned to $r=r_{\text{n}}$,
where $r_{\text{n}}$ is given by Eq. (\ref{RR}) and $w\ll D$ is a small
radial thickness of the vortex ring, while constants $\psi _{0}$ and $\chi $
are defined below. To derive an effective 1D equation, ansatz (\ref{ansatz})
is substituted in\ Eq. (\ref{GPE1}), and the standard averaging procedure is
applied, which implies the multiplication of the 2D equation by the same
localization factor which is introduced in Eq. (\ref{ansatz}), i.e., $\exp
\left( -\left( r-r_{\text{n}}\right) ^{2}/\left( 2w^{2}\right) \right) $,
followed by the integration with respect to variable $\left( r-r_{\text{n}%
}\right) $ \cite{2D-->1D}. The result is%
\begin{equation}
i\frac{\partial \psi }{\partial t}=-\frac{1}{2r_{\text{n}}^{2}}\frac{%
\partial ^{2}\psi }{\partial \tilde{\theta}^{2}}+\psi _{0}^{2}\ln \left(
\left\vert \psi \right\vert ^{2}\right) \cdot \left\vert \psi \right\vert
^{2}\psi .  \label{psi}
\end{equation}%
To cast the effective equation in this form, the constants in Eq. (\ref%
{ansatz}) are chosen as%
\begin{equation}
\psi _{0}=\exp \left( 1/\left( 8\sqrt{2}\right) \right) \approx 1.092,\chi
=1/(2w)^{2}.  \label{psi0}
\end{equation}

The CW (continuous-wave) solution of Eq. (\ref{psi}), with arbitrary
constant amplitude $a_{0}$, is%
\begin{equation}
\psi =a_{0}\exp \left[ -i\psi _{0}^{2}a_{0}^{2}\ln \left( a_{0}^{2}\right) t%
\right] .  \label{1D-CW}
\end{equation}%
Because the sign of the nonlinearity in Eq. (\ref{psi}) is self-focusing at $%
|\psi |^{2}<1$ and defocusing at $|\psi |^{2}>1$, it is relevant to analyze
the modulational instability of the CW state (\ref{1D-CW}), which is a
commonly known manifestation of the self-focusing \cite{Agrawal}. To this
end, Eq. (\ref{psi}) for the complex wave function $\psi $ is replaced by
coupled equations for the real amplitude and phase in the Madelung
representation. Next, the equations are linearized for small perturbations $%
\sim \exp \left( im\tilde{\theta}+\gamma t\right) $, where $m$ is an integer
azimuthal index of the modulational perturbation, and $\gamma $ is the
respective instability growth rate, cf. Eq. (\ref{pertur_solution}). Then, a
straightforward calculation yields
\begin{equation}
\gamma ^{2}(m)=-\frac{m^{2}}{2r_{\text{n}}^{2}}\left[ 2\psi
_{0}^{2}a_{0}^{2}\ln \left( ea_{0}^{2}\right) +\frac{m^{2}}{2r_{\text{n}}^{2}%
}\right] .  \label{gamma}
\end{equation}

The modulational instability (MI) occurs if Eq. (\ref{gamma}) gives $\gamma
^{2}>0$, the respective instability threshold being lowest for $m^{2}=1$,%
\begin{equation}
a_{0}^{2}\ln \left( \frac{1}{ea_{0}^{2}}\right) >\frac{1}{4\psi _{0}^{2}r_{%
\text{n}}^{2}}.  \label{MI}
\end{equation}%
On the contrary to the usual GP equation with cubic self-attraction, where
the MI takes place if the CW amplitude is sufficiently high, Eq. (\ref{MI})
demonstrates that, in the presence of the logarithmic factor in Eq. (\ref%
{psi}), no MI arises if the ring's radius falls below the following minimum
value:%
\begin{equation}
r_{\text{n}}\leq e/\left( 2\psi _{0}\right) \approx 1.244,  \label{no-MI}
\end{equation}%
taking $\psi _{0}$ from Eq. (\ref{psi0}).

\begin{figure}[tbp]
{\includegraphics[width=0.6\columnwidth]{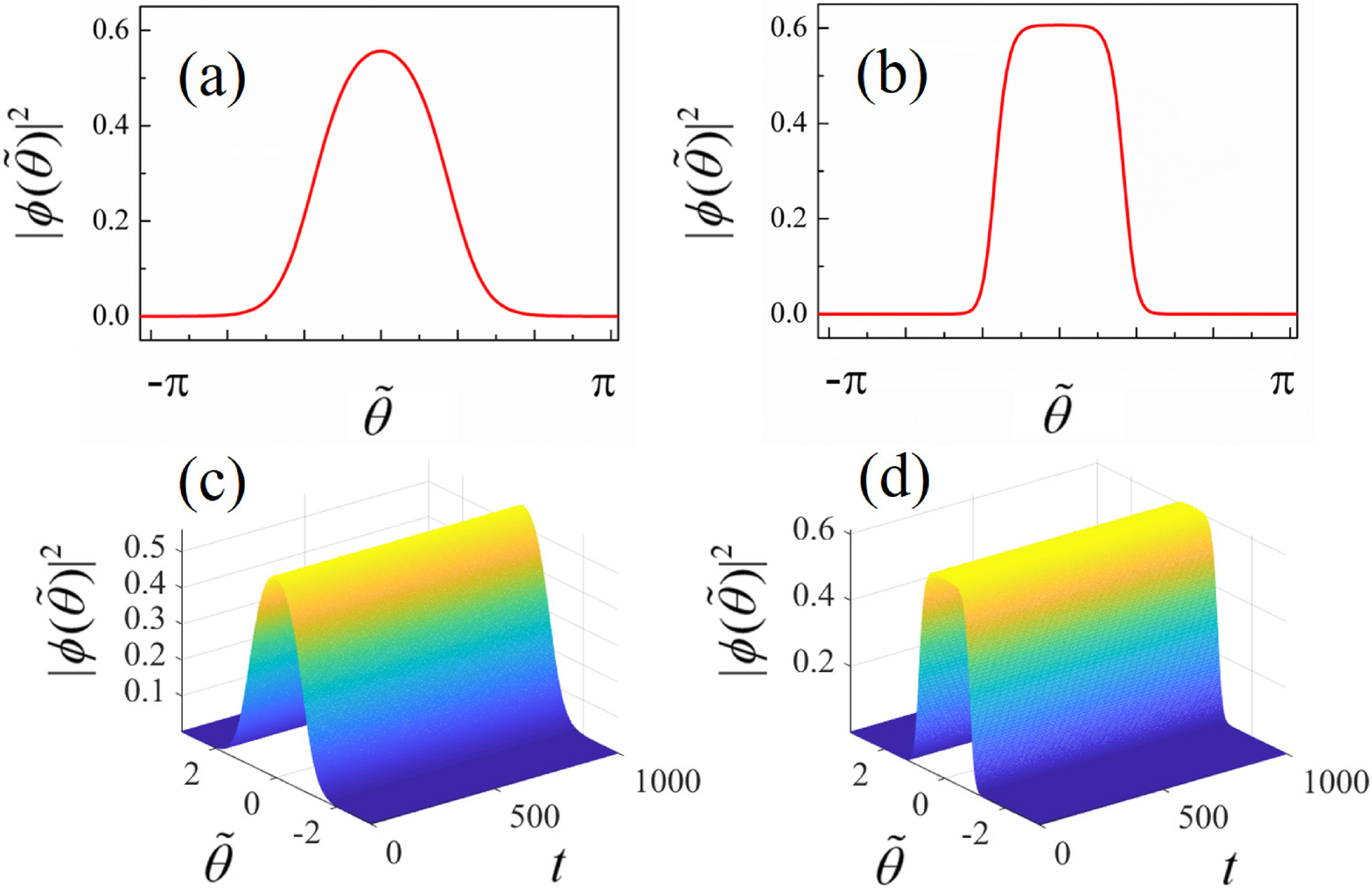}}
\caption{(Color online) Typical examples of stable 1D solitons produced by
Eq. (\protect\ref{phi}). Panels (a) and (b) display, severally, density
profiles of the solitons corresponding to $\left( N_{\mathrm{1D}}\text{,O}_{%
\text{n}}\right) =\left( 1\text{,1}\right) $ and $\left( N_{\mathrm{1D}}%
\text{,O}_{\text{n}}\right) =\left( 1\text{,2}\right) $, where $N_{\mathrm{1D%
}}$ is the 1D norm defined as per Eq. (\protect\ref{N1D}). (c,d) The
perturbed evolution of the solitons shown in panel (a) and (b), produced by
simulations of Eq. (\protect\ref{psi}) with $1\%$ random noise added to the
input. Here, the radial period is $D=6$.}
\label{soliton}
\end{figure}

If condition (\ref{MI}) holds, the MI transforms the CW into one or several
strongly localized azimuthal (quasi-1D) solitons, which may perform rotary
motion in the circular potential trough \cite{Vyslo1,Vyslo2,pre74_066615}.
In this context, soliton solutions to Eq. (\ref{psi}) with chemical
potential $\mu <0$ are looked for as%
\begin{equation}
\psi \left( \tilde{\theta},t\right) =e^{-i\mu t}\phi \left( \tilde{\theta}%
\right) ,  \label{mu}
\end{equation}%
where real function $\phi $ satisfies the equation%
\begin{equation}
\mu \phi =-\frac{1}{2r_{\text{n}}^{2}}\frac{d^{2}\phi }{d\tilde{\theta}^{2}}%
+\psi _{0}^{2}\ln \left( \phi ^{2}\right) \cdot \phi ^{3}.  \label{phi}
\end{equation}%
Figure \ref{soliton} displays typical examples of the 1D (azimuthal) soliton
solutions, obtained by means of imaginary-time-integration method with
periodic boundary conditions, which are characterized by values of the 1D
norm,%
\begin{equation}
N_{\mathrm{1D}}=\int_{0}^{2\pi }\phi ^{2}\left( \tilde{\theta}\right) d%
\tilde{\theta}.  \label{N1D}
\end{equation}%
These solutions are naturally feature spatial density profiles of two
different types, \textit{viz}., bell-shaped and flat-top ones, see Figs. \ref%
{soliton}(a) and (b), respectively. The stability of these solitons is
readily confirmed by direct simulations of their perturbed evolution, see
Figs. \ref{soliton}(c,d).

Assuming that the soliton is narrow in comparison to the range of the
variation of the angular coordinate, $2\pi $, a straightforward corollary of
Eq. (\ref{phi}) is a relation between the soliton's amplitude (maximum
value), $\phi _{0}$, and $\mu $:%
\begin{equation}
\mu =\frac{\psi _{0}^{2}}{2}\phi _{0}^{2}\left[ \ln \left( \phi
_{0}^{2}\right) -\frac{1}{2}\right] .  \label{max}
\end{equation}%
It is easy to see that Eq. (\ref{max}) admits the existence of two distinct
soliton families in a \emph{finite band} of the values of $\mu $,%
\begin{equation}
0<-\mu <\psi _{0}^{2}/\left( 2\sqrt{e}\right) \equiv -\mu _{\min }
\label{finite}
\end{equation}%
(the semi-infinite band $\mu <-\mu _{\min }$ remains empty, without any
soliton population). As $\mu $ increases from $\mu _{\min }$ to $-0$, one
soliton family has its amplitude decreasing from $\left( \phi
_{0}^{2}\right) _{\max }=1/\sqrt{e}$ to $0$. This family satisfies the
Vakhitov-Kolokolov (VK) criterion, $dN_{\mathrm{1D}}/d\tilde{\theta}<0$,
which is the well-known necessary stability condition for solitons created
by any self-focusing nonlinearity \cite{VK,Berge,Fibich}. The VK stability
of this family is corroborated, in particular, by the corresponding
dependence
\begin{equation}
N(\mu )\approx \frac{2}{r_{\text{n}}\psi _{0}^{2}}\frac{\sqrt{-2\mu }}{\ln
\left( -1/\mu \right) },  \label{N(mu)}
\end{equation}%
which is valid at $\mu \rightarrow -0$. The other, definitely unstable,
soliton family, which does not satisfy the VK criterion, features the
amplitude growing from $\phi _{0}^{2}=1/\sqrt{e}$ at $\mu =\mu _{\min }$ to $%
\phi _{0}^{2}=\sqrt{e}$ at $\mu =-0$, in the same range of the variation of $%
\mu $.

A majority of numerical results are reported below for the CW amplitudes of
the ring vortices which have $a_{0}^{2}\ln \left( 1/ea_{0}^{2}\right) <0$, hence the MI does not
occur, according to Eq. (\ref{MI}). Nevertheless, we will also consider a
nested pattern in which the inner (embedded) vortex ring is modulationally
stable, while the MI occurs in the outer ring, giving rise to an azimuthal
soliton orbiting around the inner ring. And vice versa: an azimuthal soliton
may perform rotary motion in the inner trough, which is embedded in an outer
trough filled by a modulationally stable vortex ring. Examples of such
dynamical patterns are presented below in Fig. \ref{solitons_QDs}.

\section{Numerical results}

\begin{figure}[tbp]
{\includegraphics[width=0.8\columnwidth]{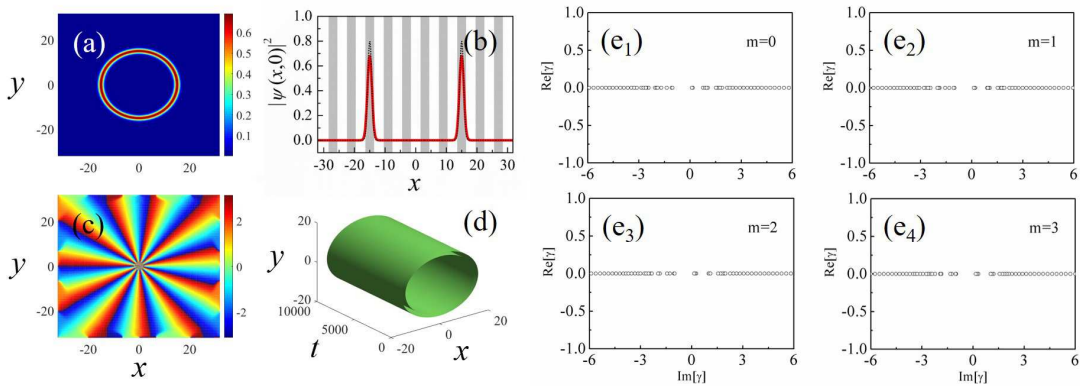}}
\caption{(Color online) (a) A stable vortex ring-shaped QD with $S=11$ and O$%
_{\text{n}}=3$. The parameters are $N=120$, $D=6$ and $V_{0}=2$. (b) Its
cross section along $y=0$, $|\protect\psi (x,0)|^{2}$, is plotted by the red
curves (the black dotted lines represent, for the comparison's sake, the
result for $V_{0}=4$, other parameters remaining the same). (c) The phase
pattern of the vortex ring. (d) Simulations of the perturbed evolution of
the vortex ring shown in panel (a) with $1\%$ random noise added to the
input. (e$_{1}$-e$_{4}$): Perturbation eigenvalues for the same vortex ring,
with azimuthal-perturbation indices $m=0,1,2,$and $3$, see Eq. (\protect\ref%
{pertur_solution}).}
\label{example}
\end{figure}

A typical example of a numerically constructed stable ring-shaped vortex QD
with $S=11$, trapped in the trough corresponding to O$_{\text{n}}=3$ in Eq. (%
\ref{RR}), is displayed in Fig. \ref{example}, the other parameters being $%
N=120$, $D=6$ and $V_{0}=2$. It corresponds to the largest WN number, $S_{%
\text{max}}$, which admits the stability for O$_{\text{n}}=3$, as shown
below in Fig. \ref{S-ON}(b). The density cross section, $|\psi (x,0)|^{2}$,
which is displayed in Fig. \ref{example}(b), corroborates the placement of
the vortex ring in the third potential trough. The stability of this QD is
confirmed by direct simulations of its perturbed evolution and spectra of
perturbation eigenvalues, as plotted in Figs. \ref{example}(d) and (e$_{1}$-e%
$_{4}$), respectively.
\begin{figure}[tbp]
{\includegraphics[width=0.7\columnwidth]{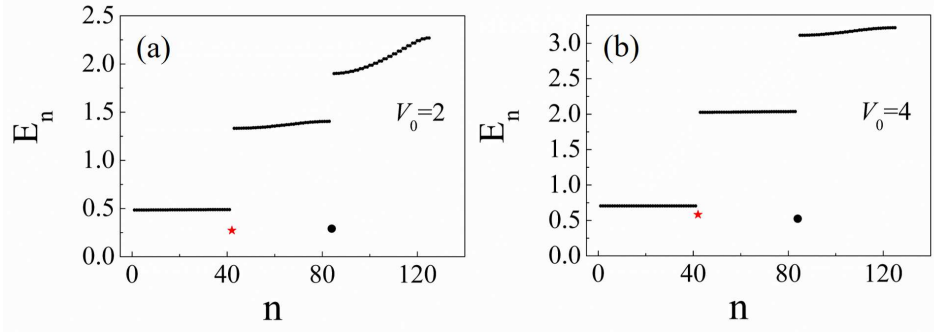}}
\caption{(Color online) The black curves represent the energy levels $%
\mathrm{E}_{\mathrm{n}}$ (eigenvalues of chemical potentials $\protect\mu $)
corresponding to the lowest bands representing delocalized radial
eigenstates produced by numerical solution of the linearized version of Eq. (%
\protect\ref{radial}) with $V_{0}=2$ in (a) and $V_{0}=4$ in (b) in the
limit of $r\rightarrow \infty $. Here $\mathrm{n}$, attached to the
horizontal axes, is the index of the eigenstates. Symbols $\bigstar $\emph{\
}and\emph{\ $\bullet $ }in (a) designate, respectively, values of $\protect%
\mu $ for stable vortex-ring QDs with $S=3$, O$_{\text{n}}=3$, $N=160$, and $%
S=2$, O$_{\text{n}}=2$, $N=100$. In panel (b), the same symbols represent
the ring-shaped QDs with $S=1$, O$_{\text{n}}=1$, $N=30$, and $S=4$, O$_{%
\text{n}}=4$, $N=200$, respectively. Here, the radial period of potential (%
\protect\ref{potential}) is $D=6$. }
\label{gap}
\end{figure}

It is relevant to calculate the bandgap spectrum of the linearized version
of Eq. (\ref{radial}) with $r\rightarrow \infty $, and identify the position
of the QD's eigenvalues $\mu $ with respect to the spectrum. Typical
examples demonstrate in Fig. \ref{gap}, for $V_{0}=2$ and $4$ [see Eq. (\ref%
{potential})], that the eigenvalues belong to the semi-infinite bandgap,
located beneath all Bloch bands populated by delocalized radial eigenmodes.
The same conclusion was made for other stable ring-shaped states.

\begin{figure}[tbp]
{\includegraphics[width=0.8\columnwidth]{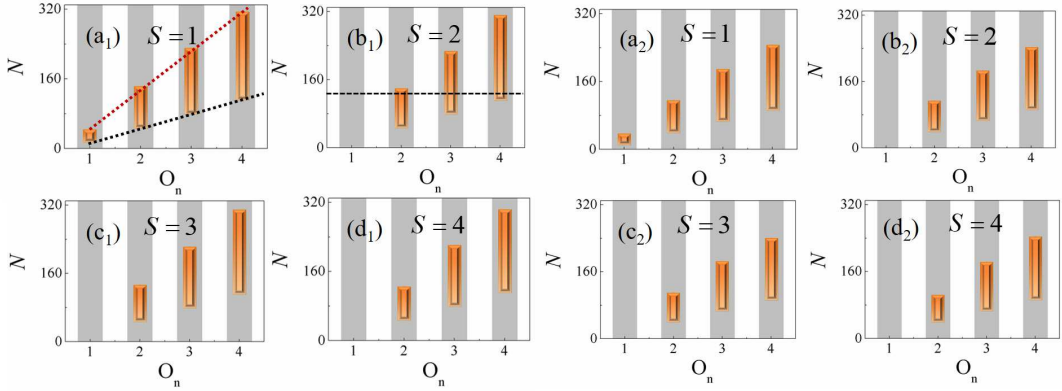}}
\caption{(Color online) Orange bars represent stability intervals for the
vortex ring-shaped QDs with $S=1,2,3$ and $4$, which are placed in the
radial-potential troughs with O$_{\text{n}}=1,2,3$ and $4$ [see Eq. (\protect
\ref{RR})]. The red and black oblique dotted lines in panel (a$_{1}$) show
the expansion of the stability area with the increase of O$_{\text{n}}$. The
horizontal dashed line in panel (b$_{1}$) highlights the multistability, in
the form of the coexistence of the vortex rings with the same norm but
different values of O$_{\text{n}}$. The depth of the radial potential (%
\protect\ref{potential}) is $V_{0}=2$ in panels (a$_{1}$-d$_{1}$) and $%
V_{0}=4$ in (a$_{2}$-d$_{2}$), while its radial period is $D=6$.}
\label{Change_N_n_V=2+4D=6_S=1234}
\end{figure}

\begin{figure}[tbp]
{\includegraphics[width=0.6\columnwidth]{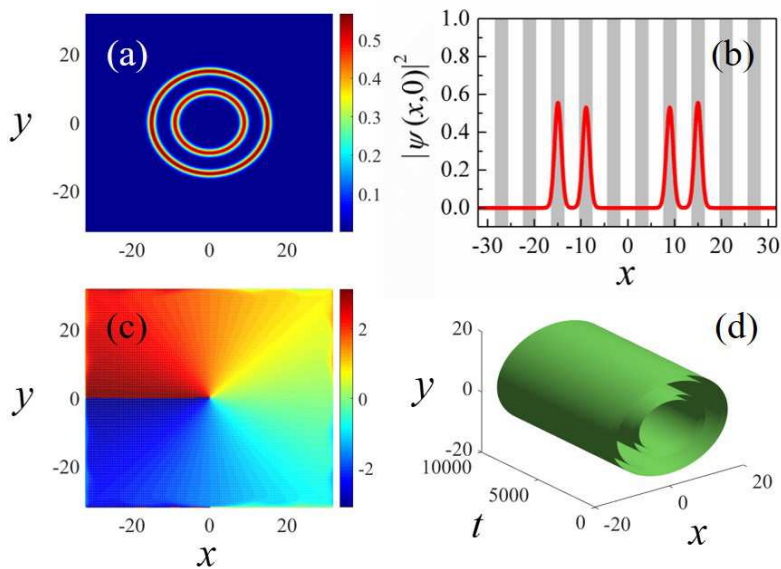}}
\caption{(Color online) (a) An example of a stable ring-shaped vortex QDs
with $S=1$, which fills two radial troughs, with O$_{\text{n}}=2$ and $3$.
(b) Its cross section, $|\protect\psi (x,0)|^{2}$, along $y=0$. (c) The
corresponding phase patterns. (d) Direct simulations of the perturbed
evolution of the two-ring vortex QDs shown in panel (a) with $1\%$ random
noise added to the input. Here, the norm is $N=150$, and other parameters
are the same as in Fig. (\protect\ref{example}).}
\label{N=150_V=2_D=6_ON23}
\end{figure}

Results of the numerical analysis of the stability of the vortex-ring QDs,
with different WNs $S$, are summarized in Fig. \ref%
{Change_N_n_V=2+4D=6_S=1234} for different values of the radial-potential's
depth $V_{0}$ [see Eq. (\ref{potential})]. In the figure, the stability
areas are shown by orange bars in the plane of norm $N$ and radial-trough's
number O$_{\text{n}}$, see Eq. (\ref{RR}). It is seen that,\ for fixed
values of $S$ and $V_{0}$, the vortex rings are stable in a finite interval,
\begin{equation}
N_{\min }<N<N_{\max }.  \label{minmax}
\end{equation}%
At $N<N_{\min }$, the vortex QDs with the given parameters do not exist,
while at $N>N_{\max }$ the peak (largest) density of the QDs essentially
exceeds characteristic values $n\sim 1$ determined by the nonlinearity in
Eq. (\ref{GPE1}), which leads to overfill of the given radial trough and
filling adjacent troughs [see an examples in Fig. \ref{N=150_V=2_D=6_ON23},
in which the condensate fills two troughs]. The nonexistence of the QDs at $%
N<N_{\min }$ is explained by the fact that, for small values of $N$, the
density falls to a value close to $n=1$, at which the effective nonlinearity
in Eq. (\ref{GPE1}), containing the logarithmic factor, vanishes, while the
linear version of Eq. (\ref{GPE1}) with potential (\ref{potential}) cannot
support any state localized in the radial direction.

\begin{figure}[tbp]
{\includegraphics[width=0.7\columnwidth]{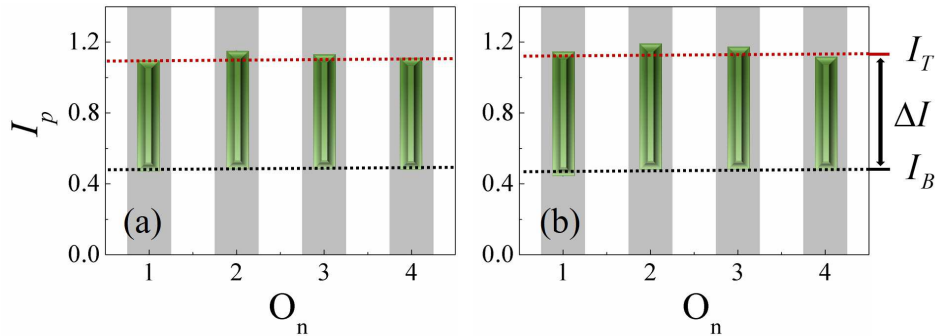}}
\caption{(Color online) The peak density of the stable vortex ring-shaped
QDs with $S=1$, which are placed in the radial troughs with O$_{\text{n}%
}=1,2,3$ and $4$. The parameters are $D=6$ and $V_{0}=2$ in (a), or $V_{0}=4$
in (b). The numerical results yield $I_{B}\approx 0.48$, $I_{T}\approx 1.14$%
, and $\Delta I=I_{T}-I_{B}\approx 0.66$.}
\label{density}
\end{figure}

Note that the length of the stability interval in Fig. \ref%
{Change_N_n_V=2+4D=6_S=1234} increases with the growth of the trough's
number, O$_{\text{n}}$. This trend can be explained as follows: according to
Refs. \cite{FOP_ZYY,CSF152_111313}, vortex QDs are stable if their peak
density takes values in a finite range around $n\sim 1$. To address this
point in detail, green bars in Fig. \ref{density} designate the peak density
($I_{p}$) of the stable vortex QDs with $S=1$, which are placed in the
radial troughs with numbers O$_{\text{n}}=1,2,3$ and $4$. It is observed
that $I_{p}$ ranges between the bottom and top values, $I_{B}$ and $I_{T}$
(the black and red dotted lines, respectively, in Fig. \ref{density}).
Further, the length of the narrow vortex ring is $l_{n}=2\pi r_{\text{n}%
}=\pi (2$O$_{\text{n}}-1)D$ [see Eq. (\ref{RR})]. Thus, the boundaries of
the stability interval (\ref{minmax}) can be written as
\begin{equation}
\left\{ N_{\min },N_{\max }\right\} =\pi \left\{ I_{B},I_{T}\right\} (2\text{%
O}_{\text{n}}-1)Dw/2,  \label{NN}
\end{equation}%
where $w$, which depends on $V_{0}$ and $D$, is an effective width of the
narrow QD in the radial direction, cf. Eq. (\ref{ansatz}). The respective
width of the stability interval in Fig. \ref{Change_N_n_V=2+4D=6_S=1234} is
\begin{equation}
\Delta N=N_{\max }-N_{\min }=(\pi /2)\Delta I\cdot (2\text{O}_{\text{n}%
}-1)Dw,  \label{LN}
\end{equation}%
where $\Delta I=I_{T}-I_{B}$. Thus, the linear expansion of the stability
interval with the increase of O$_{\text{n}}$ is explained by Eq. (\ref{LN}).

Comparing the above results for $V_{0}=2$ [Figs. \ref%
{Change_N_n_V=2+4D=6_S=1234}(a$_{1}$-d$_{1}$)] and $V_{0}=4$ [Figs. \ref%
{Change_N_n_V=2+4D=6_S=1234}(a$_{2}$-d$_{2}$)], we conclude that the
stability intervals are slightly longer in the former case. This finding can
be explained too: the increase of the modulation depth of the potential, $%
V_{0}$, causes slight decrease of the radial width $w$ of the vortex rings
[see the cross section $|\psi (x,0)|^{2}$ in Fig. \ref{example}(b), where
the red and black curves represent $V_{0}=2$ and $V_{0}=4$, respectively].
According to Eq. (\ref{LN}), this results in a decrease in $\Delta N$.

If $V_{0}$ is fixed, the results reported in Ref. \cite{CNSNS_LZD}
demonstrate that the ring-shaped vortex QDs with a large topological charge
tend to require a larger radius. For this reason, as shown in Fig. \ref%
{Change_N_n_V=2+4D=6_S=1234}(b$_{1}$), the vortex ring with $S=2$ cannot
stably exist in the innermost radial trough corresponding to O$_{\text{n}}=1$
in Eq. (\ref{RR}).

The present setting admits multistability of the vortex-ring QDs, as shown,
in particular, by the horizontal dashed line in Fig. \ref%
{Change_N_n_V=2+4D=6_S=1234}(b$_{1}$)]. It highlights the coexistence of the
stable vortex rings with equal norms, hosted by the radial troughs with
different values of O$_{\text{n}}$ but the same modulation depths.

Stable vortex QDs which fill more than one trough exist as well, see an
example for $S=1$ in Fig. \ref{N=150_V=2_D=6_ON23}. Actually, solutions of
this type may be considered as an example of the concentric (nested)
multi-ring vortex QDs, in the particular case when the coupled rings carry
the same topological charge. Nested multi-ring complexes are considered in
detail in the next section.

\begin{figure}[tbp]
{\includegraphics[width=0.75\columnwidth]{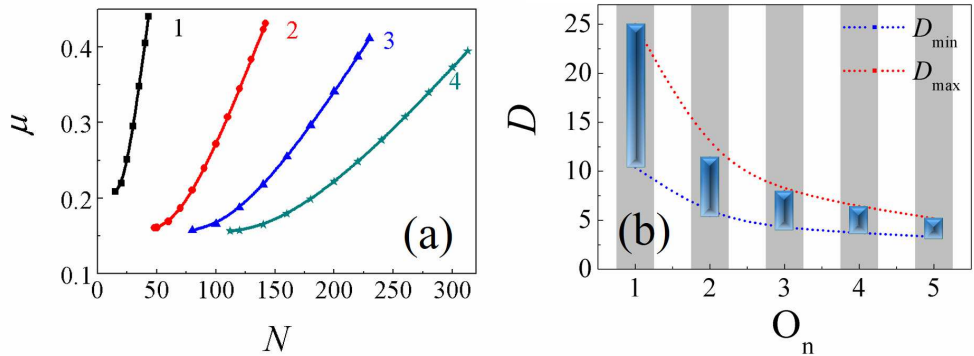}}
\caption{(Color online) (a) The dependence of the chemical potential, $%
\protect\mu $, on norm, $N$, for the ring-shaped vortex QDs, which populate
the radial troughs with O$_{\text{n}}=1,2,3$ and $4$, respectively. The
parameters are the same as in Fig. \protect\ref{Change_N_n_V=2+4D=6_S=1234}(a%
$_{1}$). (b) The boundary values, $D_{\min }$ and $D_{\max }$, of the period
of the radial potential (\protect\ref{potential}), between which the vortex
rings are stable, for given values of O$_{\text{n}}$ [see Eq. (\protect\ref%
{DDD})]. The other parameters are $N=120$, $V_{0}=2$, and $S=1.$}
\label{MU+D}
\end{figure}

The dependence of the chemical potential, $\mu $, of the ring-shaped vortex
QDs on the total norm, $N$, is plotted in Fig. \ref{MU+D}(a). The $\mu (N)$
curves in Fig. \ref{MU+D}(a) feature a positive slope, $d\mu /dN>0$, which
contradicts the VK criterion. On the other hand, in the case of the
effective self-repulsion in Eq. (\ref{GPE1}), which is the case for these
solutions, the necessary stability condition for localized states is the
\textit{anti-VK} criterion, which is exactly $d\mu /dN>0$ \cite{anti}. In
the case of the sign-changing nonlinearity, such as that in Eq. (\ref{GPE1}%
), or the combination of cubic self-focusing and quintic defocusing in 1D
\cite{1D_AVK_CQ} and 2D \cite{checkerboard_potential,FOP_ZYY} models, the
necessary sign of the slope corresponding to stable localized states
switches between $d\mu /dN<0$ and $d\mu /dN>0$.

Here, we provide an analysis to explain why the $\mu (N)$ curves for the
vortex ring-shaped QDs obey the anti-VK criterion. We assume the value of
the peak density of QDs is $I_{p}$, which can be estimated by $I_{p}\approx
N/[\pi (2$O$_{\text{n}}-1)Dw/2]$. According to Ref. \cite{FOP_ZYY}, when $%
N>N_{\mathrm{threshold}}=\pi (2$O$_{\text{n}}-1)Dw/(2e)$, the anti-VK
criterion, $d\mu /dN>0$, is relevant as the necessary stability condition.
In Fig. \ref{MU+D}(a), the radial period is fixed as $D=6$. If we select $%
w=D/2$, $N_{\mathrm{threshold}}$ for O$_{\text{n}}=1,2,3$ and $4$ are $N_{%
\mathrm{threshold}}\approx 10.4,31.2,52.0$ and $72.8$, respectively. As
shown in Fig. \ref{MU+D}(a), the numerically found bottom stability
boundaries for O$_{\text{n}}=1,2,3$ and $4$ are of $N_{\min }\approx
15,48,80 $ and $112$, respectively, which all are larger than $N_{\mathrm{%
threshold}}$ for each value of O$_{\text{n}}$, hence the anti-VK criterion
determines the stability in this case.

Multistability is also revealed by $\mu (N)$ curves Fig. \ref{MU+D}(a). It
is realized as the coexistence of vortex rings with equal values of the
norm, trapped in the radial troughs with different values of O$_{\text{n}}$
but the same potential depth $V_{0}$.

The effect of the period of the radial potential, $D$, on the ring-shaped
vortex QDs is presented in Fig. \ref{MU+D}(b), where $N=120$, $V_{0}=2$, and
$S=1$ are fixed. In this figure, the vortex ring are stable in intervals%
\begin{equation}
D_{\min }<D<D_{\max },  \label{DDD}
\end{equation}%
at given values of O$_{\text{n}}$. The fact that the length of the stability
interval (\ref{DDD}) decreases with the increase of O$_{\text{n}}$ can be
explained by dint of Eq. (\ref{NN}), which, for fixed $N$, yields
\begin{equation}
\Delta D\equiv D_{\max }-D_{\min }=\frac{N\Delta I}{\pi w(\text{O}_{\text{n}%
}-1/2)I_{B}I_{T}}.  \label{DeltaD}
\end{equation}%
Note that the inverse proportionality of $\Delta D$ to $($O$_{\text{n}}-1/2)$%
, demonstrated by Eq. (\ref{DeltaD}), agrees with the numerical findings
presented in Fig. \ref{MU+D}(b).

\begin{figure}[tbp]
{\includegraphics[width=0.75\columnwidth]{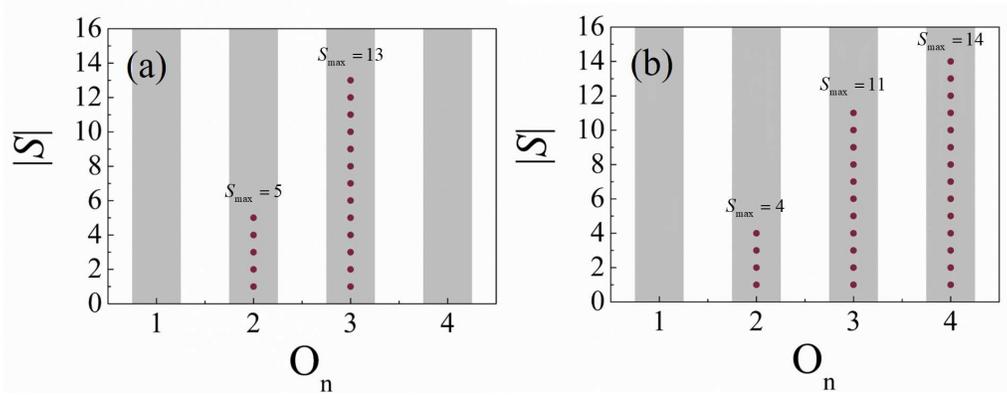}}
\caption{(Color online) The trapping ability of the radial-potential troughs
for the vortex rings with WN\ (winding number) $\left\vert S\right\vert $
versus O$_{\text{n}}$: (a) $N=100$ and (b) $N=120$. The other parameters are
$D=6$ and $V_{0}=2$.}
\label{S-ON}
\end{figure}

The ability of the radial potential (\ref{potential}) to maintain vortex
rings with high values of $S$ at different values of O$_{\text{n}}$, as
produced by the numerical solution, is summarized in Figs. \ref{S-ON}(a) and
(b) for $N=100$ and $N=120$, respectively, while the other parameters are
fixed, $D=6$ and $V_{0}=2$. The results are consistent with those presented
in Fig. \ref{Change_N_n_V=2+4D=6_S=1234}(a$_{1}$). For these values of $N$,
no stable solutions are found at O$_{\text{n}}=1$, for instance, $N_{\max
}=43$ for $S=1$ at O$_{\text{n}}=1$, cf. Eq. \ref{N}. Similarly, no stable
vortex rings with $N=100$ exist in O$_{\text{n}}=4$, because $N_{\min }=112$
for $S=1$ at O$_{\text{n}}=4$. Generally, Fig. \ref{S-ON} shows that the
holding capacity of the radial troughs increases with the growth of O$_{%
\text{n}}$. Comparison between Figs. \ref{S-ON}(a) and (b) also reveals that
the increase of the QD's norm reduces the capacity of the trough with given O%
$_{\text{n}}$.

\begin{figure}[tbp]
{\includegraphics[width=0.8\columnwidth]{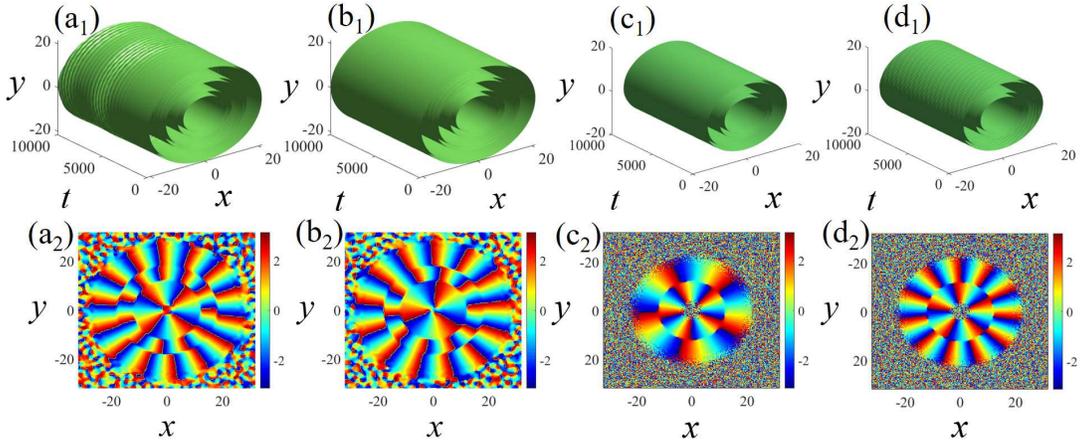}}
\caption{(Color online) Typical examples of stable nested (multi-ring)
vortex QDs. (a$_{1}$) A set of three rings characterized by $(N,S$,O$_{\text{%
n}})$ $=$ $(120,4,2)$, $(120,11,3)$, and $(N,S$,O$_{\text{n}})$ $=$ $%
(120,14,4)$. (b$_{1}$) The same, but with $(N,S$,O$_{\text{n}})$ $=$ $%
(120,3,2)$, $(120,10,3)$, and $(N,S$,O$_{\text{n}})$ $=$ $(120,13,4)$. (c$%
_{1}$) and (d$_{1}$) Typical examples of stable concentric two-ring vortex
QDs with opposite signs of $S$ in the coupled rings, whose parameters are ($%
N $,$S$,O$_{\text{n}})$ $=$ $(100,5,2)$ and $(100,-5,3)$, or $(100,5,2)$ and
$(100,-11,3)$, respectively. (a$_{2}$-d$_{2}$) Output phase patterns of the
corresponding three- and two-ring states at $t=10000$, which corroborates
their structural stability. The other parameters are the same as in Fig.
\protect\ref{S-ON}.}
\label{nest_QDs}
\end{figure}

\begin{figure}[tbp]
{\includegraphics[width=0.7\columnwidth]{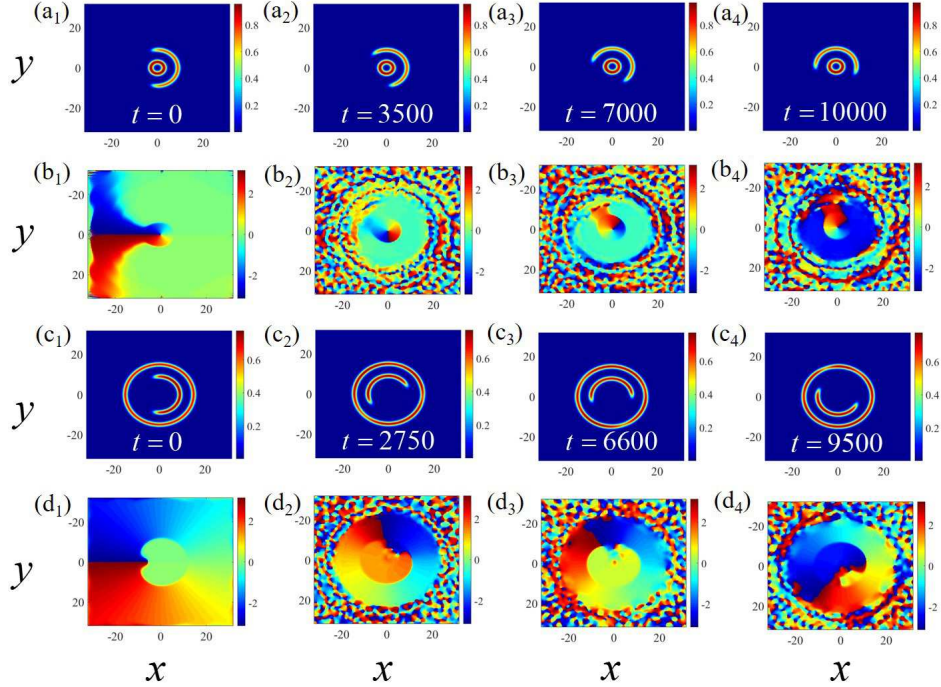}}
\caption{(Color online) Typical examples of stable nested patterns with
soliton and vortex QDs. which were created in adjacent radial troughs. In
panels (a$_{1}$-b$_{4}$) the pattern was created from the initial dynamical
states with parameters $(N$,$S$,O$_{\text{n}})$ $=$ $(46,0,2)$ and $(N$,$S$,O%
$_{\text{n}})$ $=$ $(35,1,1)$ in the outer and inner troughs, respectively.
In panels (c$_{1}$-d$_{4}$) the input was taken with parameter sets $(N$,$S$%
,O$_{\text{n}})$ $=$ $(120,1,3)$ and $(N$,$S$,O$_{\text{n}})=$ $(46,0,2)$ in
the outer and inner troughs.}
\label{solitons_QDs}
\end{figure}

\section{Multi-ring (nested) vortex quantum droplets}

Typical examples of concentric multi-ring (nested) states are displayed in
Fig. \ref{nest_QDs}. Most interesting ones are the states composed of
individual vortex rings with \emph{different} WNs, cf. Ref. \cite{Skarka}.
As generic examples of such states, Figs. \ref{nest_QDs}(a$_{1}$) and (b$%
_{1} $) demonstrate robust three-ring solutions built of vortex rings with,
respectively, ($N$,$S$,O$_{\text{n}})$ $=$ $(120,4,2)$, $(120,11,3)$, and $%
(120,14,4)$, or $(120,3,2)$, $(120,10,3)$, and $(120,13,4)$. In direct
simulations, such compound states remain stable for times exceeding $t=10000$%
, which corresponds to $\sim 300$ diffraction times. These results offer a
use in the design of encoding devices which employ winding numbers for the
data storage.

It is worthy to note that it is also possible to construct stable concentric
two-ring modes with \emph{opposite signs} of $S$ in the rings. This is
somewhat similar to states with \textit{hidden vorticity}, which are
produced by a system of 2D GP equations including the self- and
cross-component cubic nonlinear terms and the isotropic harmonic-oscillator
potential trapping each component \cite{Brtka,PRA_LYY_2DQD2018}. Such states
with vorticities $S=\pm 1$ in the two components have their stability region
in the underlying parameter space. Typical examples of the stable concentric
two-ring vortex QDs of this type, with parameters of the coupled rings ($N$,$%
S$,O$_{\text{n}})$ $=$ $(100,5,2)$ and $(100,-5,3)$, or $(100,5,2)$ and $%
(100,-11,3)$, are displayed in Fig. \ref{nest_QDs}(c$_{1}$) and (d$_{1}$),
respectively. To confirm the stability of the three- and two-ring compound
modes, Figs. \ref{nest_QDs}(a$_{2}$-d$_{2}$) depict the respective output
phase patterns at $t=10000$.

As mentioned above, another noteworthy option is to construct a two-ring
complex in which one vortex-ring component is subject to the MI, hence it is
replaced by an azimuthal soliton (or maybe several solitons), see Eqs. (\ref%
{mu}) and (\ref{phi}), while the vortex component trapped in another
potential trough avoids the azimuthal MI and remains essentially
axisymmetric. Examples of such heterogeneous robust states, produced by
simulations of Eq. (\ref{GPE1}), are displayed in Fig. \ref{solitons_QDs}.
Panels \ref{solitons_QDs}(a$_{1}$-b$_{4}${) show a complex in which the MI
takes place in the outer circular trough, producing an azimuthal soliton
which performs rotary motion, while the inner vortex ring is modulationally
stable. An opposite example is produced in Figs. \ref{solitons_QDs}(}c$_{1}$%
-d$_{4}${), where the outer vortex ring remains stable against azimuthal
perturbations, while the MI creates a soliton exhibiting the rotary motion
in the embedded (inner) circular trough. The rotation direction of the
soliton is driven by the vorticity sign of the underlying QD. }

It is relevant to mention that the multi-ring potential considered here
holds different vortex-ring or azimuthal-soliton states nearly isolating
them from each other. In particular, this property offers the
above-mentioned potential for the design of the data-storage device, in
which different data components may be kept in different radial troughs. An
additional problem, which is left for subsequent analysis, is interplay
between adjacent radial modes in the case when the separation between the
adjacent rings is essentially smaller.

\section{Conclusion}

The subject of this work is the class of vortical ring-shaped QDs (quantum
droplets) trapped in the 2D radially periodic potential. The underlying
model is the GP equation with the cubic nonlinearity multiplied by the
logarithmic factor, which represents the correction to the mean-field theory
produced by quantum fluctuations. Effects of the depth and period of the
potential on the vortex rings are studied, and their stability area is
identified. Multistability of the vortex QDs is addressed too, demonstrating
the coexistence of narrow vortex rings with equal norms in the
radial-potential troughs with different radii. For the narrow rings, the 2D
GP\ equation is approximately reduced to the 1D equation, which makes it
possible to analyze the modulational stability of the rings against
azimuthal perturbations. Multi-ring states in the form of nested patterns
with different topological charges trapped in different radial troughs are
found too. Also produced are robust dynamical states with modulationally
stable narrow ring trapped in a particular circular trough, and an azimuthal
soliton, created by the azimuthal modulational instability, in an adjacent
trough. The results reported in this work suggest a new approach to the
creation of stable QDs with embedded vorticity, which is a challenging
problem. The results may also be used in the design of data-storage devices,
with stable vortex rings trapped in different radial troughs used for
encoding separate components of the data.

The present analysis can be extended in other directions. In particular, it
may be interesting to generalize the model by considering an elliptically
deformed radial potential. A challenging possibility is to the extend the
present analysis for three-dimensional settings.

\section{Acknowledgments}

This work was supported by NNSFC (China) through grant Nos. 11905032,
11874112, 11904051, Natural Science Foundation of Guangdong province through
grant No. 2021A1515010214, the Guang Dong Basic and Applied Basic Research
Foundation through grant No. 2021A1515111015, the Key Research Projects of
General Colleges in Guangdong Province through grant No. 2019KZDXM001, the
Special Funds for the Cultivation of Guangdong College Students Scientific
and Technological Innovation through grant Nos. pdjh2021b0529,
pdjh2022a0538, the Research Fund of Guangdong-Hong Kong-Macao Joint
Laboratory for Intelligent Micro-Nano Optoelectronic Technology through
grant No. 2020B1212030010, and Israel Science Foundation through grant No.
1695/22.

\end{document}